# Super-Resolution Imaging by Arrays of High-Index Spheres Embedded in Transparent Matrices

Kenneth W. Allen,[1,2] Navid Farahi,[1,2] Yangcheng Li,[1] Nicholaos I. Limberopoulos,[2] Dennis E. Walker Jr.,[2] Augustine M. Urbas,[3] and Vasily N. Astratov[1,2*]

[1]Department of Physics and Optical Science, Center for Optoelectronics and Optical Communications, University of North Carolina at Charlotte, 9201 University City Blvd., Charlotte, NC 28223-0001, USA
[2]Sensors Directorate, United States Air Force Research Laboratory, Wright-Patterson AFB, OH 45433, USA
[3]Materials and Manufacturing Directorate, United States Air Force Research Laboratory, Wright-Patterson AFB, OH 45433, USA
*Tel: (704) 687 8131, Fax: 1 (704) 687 8197, Email: {kallen62, astratov}@uncc.edu

*Abstract*— We fabricated thin-films made from polydimethylsiloxane (PDMS) with embedded high-index ($n$~1.9-2.2) microspheres for super-resolution imaging applications. To control the position of microspheres, such films can be translated along the surface of the nanoplasmonic structure to be imaged. Microsphere-assisted imaging, through these matrices, provided lateral resolution of ~$\lambda$/7 in nanoplasmonic dimer arrays with an illuminating wavelength $\lambda$=405 nm. Such thin films can be used as contact optical components to boost the resolution capability of conventional microscopes.

## I. INTRODUCTION

In recent years imaging by dielectric microspheres appeared as a paradigm shift in microscopy [1-7]. As illustrated in Fig. 1, this technique is extremely simple. It uses a dielectric microsphere as a "magnifying lens" creating a virtual image of an object located at the surface of the structure. Visualization of the virtual images of the surface requires a change of the depth of focusing, as schematically illustrated in Fig. 1. According to classical diffraction limit the resolution of such a system should be determined by $d=\lambda/(2n\sin\theta)$, where $\lambda$ is the wavelength of the illuminating source, $n$ is the refractive index of the object space and $\theta$ is the half-angle over which the objective can accept light from the object. By assuming that we use high-index spheres with $n$~2.0 in contact with the object and that the light is collected from a semi-space ($\sin\theta$~1), one can estimate the maximal far-field resolution as $d=\lambda/4$. It should be noted, however, that significantly better resolution has been reported by many groups [1-7]. These studies have been performed on different objects including plasmonic nanostructures and biological objects. Although different groups used various definitions of resolution, it seems that according to a rigorous textbook definition, the resolution of at least $d=\lambda/7$ has been demonstrated by this method [6]. This situation creates many questions about the mechanisms of such microsphere-assisted imaging. Roles of surface excitations, plasmons, polaritons and Tamm-states, as well as the role of nanoscale gap separating the object and lens have been intensively discussed in the literature.

Fundamentally, a connection between the super-resolution strength of the microspheres and their ability to focus light into sub-diffraction limited dimensions has also been noted [1]. These tightly focused beams have been termed "photonic-nanojets" [8-10] and corresponding "nanojet-induced modes" [11-19] have experimentally been observed in single spheres and chains of spheres, respectively.

To develop practical applications of this method in microscopy it is important to find ways to precisely align the spheres with the object of study. It can be achieved by individual manipulation of microspheres using positioning tools, termed locomotion of microspheres [5]. This procedure, however, is rather slow and complicated if different spheres with various sizes and indices need to be tested for imaging.

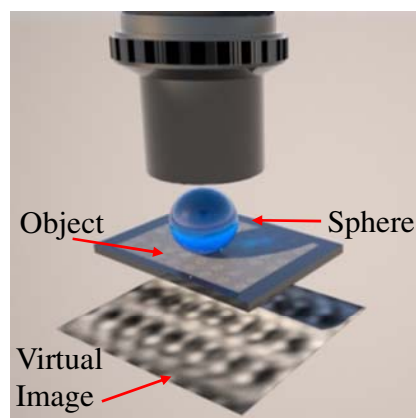

Fig. 1: Schematic of the microsphere-assisted imaging technique. A microsphere is placed on top of an object, and a virtual image is projected into the far-field. This virtual image is then collected by the objective lens.

In this work, we developed a different approach to this problem based on embedding high-index microspheres in a liquid material with an ability to solidify. In our studies we used polydimethylsiloxane (PDMS) for embedding the microspheres. Similar technology has been developed previously for using microspheres as lenses for projection lithography [20] and as photonic circuits [21]. Once such transparent matrices with embedded microspheres are hardened, they can be applied to the surface of the investigated structure and mechanically translated along the surface to align different spheres with the object of interest. We show that lubrication of the surface with isopropanol (IPA) provides good conditions for the lateral translation of such matrices with embedded spheres.

## II. EXPERIMENT

In order to fabricate such arrays we implemented a three-step process, illustrated in Fig. 2(a-c). First, the two-dimensional array of high-index barium titanate glass (BTG) microspheres (Mo-Sci Corp.), which vary in diameter ($9 < D < 212$ $\mu$m), was deposited onto a substrate, see Fig. 2(a). We used two types of BTG spheres with indices 1.9 and 2.1, respectively. Then, a PDMS (Dow Corning Corp.) layer was cast over the spheres and cured in a temperature controlled oven, as shown in Fig. 2(b). Finally, after the PDMS layer had cured into an elastomeric state from the thermal treatment, the thin-film was lifted from the substrate by a scalpel, as shown in Fig. 2(c).

We performed structural characterization of the thin-films to demonstrate that the spheres protrude from the bottom surface. This is important for super-resolution imaging since the spheres should be placed close to the object being imaged. The lower surface was found to be sufficiently smooth for developing applications. The analysis of the surface performed by atomic force microscopy showed that the surface relief experiences ~200 nm variations making bumps where the embedded spheres are reaching the bottom surface. Thus, the spheres are found to be in a nanoscale vicinity of the lower surface of matrices. A droplet of liquid (IPA) was applied between the investigated sample and the PDMS thin film to simplify the translation of the thin films by reducing friction.

The thin-film was applied and scanned across the investigated surface and an image was obtained using the embedded microspheres. We employed two imaging systems to capture virtual images produced by the microspheres: i) The FS70 Mitutoyo microscope equipped with a halogen lamp, that has a strong spectral response peaked at $\lambda$ ~550 nm, and a CCD camera operated in reflection illumination mode; ii) Scanning laser confocal microscope (SLCM) Olympus LEXT-OLS4000 at $\lambda$ = 405nm. The examples of imaging BTG spheres embedded in PDMS matrices by the former system are presented in Figs. 2(d) and 2(e). In these images the depth of focus was close to the equator plane of spheres.

The advantage of the fabricated thin-films, in comparison with previous studies of fully liquid-immersed spheres [3,6], is based on the ability to translate the entire thin-film along the surface, which allows very simple alignment of different spheres with our object of study. A metallic probe, connected to a hydraulic micromanipulation controller was pressed into the thin-film providing its lateral translation. The widefield microscopy allowed for precise control of the alignment of different spheres with different surface nanostructures.

## III. RESULTS AND DISCUSSION

As an object of study, we used the nanoplasmonic arrays of dimers fabricated by e-beam lithography at the Air Force Research Laboratory. The dimers had varying diameters of golden cylinders and edge-to-edge separations. We present results obtained for dimers with diameter of 175 nm and with center-to-center distance 200 nm, leading to a 25 nm (~$\lambda$/16) edge-to-edge separation, as illustrated by SEM image in Fig. 3(a). These arrays were deposited on a sapphire substrate, first with 10 nm of titanium and followed by 40 nm of gold (Au). The imaging was provided through a ~13 $\mu$m BTG sphere embedded in PDMS, captured by a scanning laser confocal microscope with a 100×(NA=0.95) objective at $\lambda$=405 laser illumination, as illustrated in Fig. 3(c).

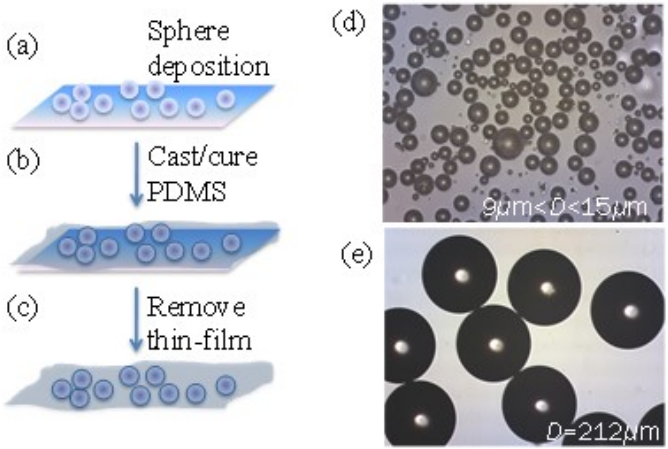

Fig. 2: (a-c) Illustration of the fabrication process for transparent PDMS matrices with high-index spheres embedded, (d) Optical microscope image of BTG spheres with diameters from 9 to 15 $\mu$m embedded in a PDMS matrix, (e) Optical microscope image of BTG spheres with diameters 212 $\mu$m embedded in a PDMS matrix.

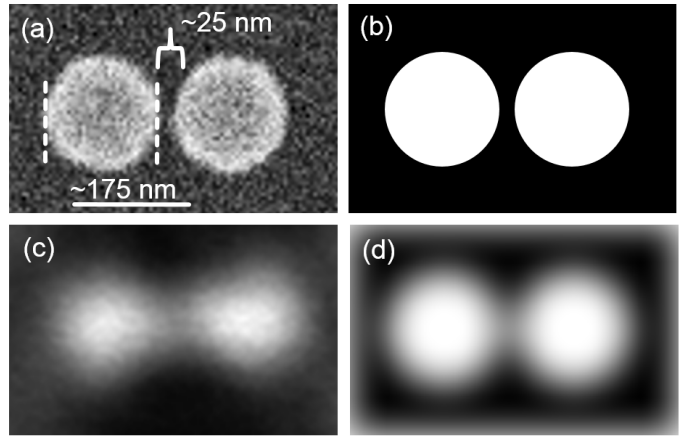

Fig. 3: (a) Scanning electron microscope image of the gold nanoplasmonic dimer. (b) Drawn object used for calculating a convoluted image. (c) Image of the dimer obtained through the microsphere embedded in the PDMS thin-film. (d) Calculated image, showing the convolution of the drawn object in (b) with a point spread function with the width of ~ $\lambda$/7.

The resolution was determined by using a convolution theorem, which states that an image is the result of the convolution of an object with the imaging system point spread function (PSF). We convoluted drawn object in Fig. 3(b) with the Gaussian PSFs with different widths in order to compare the results with the experimental image. Good agreement with the experiment was obtained for the PSF width of $\lambda/7$, shown in Fig. 3(d). Thus, we estimate the resolution, calculated as the width of PSF according to Houston's criterion, on the order of $\lambda/7$.

## IV. SUMMARY

In this work, we designed and tested arrays of high-index BTG spheres embedded in transparent PDMS matrices. We show that such thin films containing spheres are a unique tool for super-resolution microscopy since different spheres can be easily aligned with the object of study by shifting the thin-film along the surface. Many questions about the mechanism of super-resolution imaging by microspheres still require further investigation. These include the role of the nanometric gap between the object and spheres and the role of surface polariton-plasmons in the metallic nanostructures. Our results, however, show that spheres embedded in elastomeric PDMS matrices can behave in these applications similar to liquid-immersed spheres studied previously [3,6]. Due to resolution on the order of $\lambda/7$, excellent manufacturability of such thin films and their easy-of-use, such optical components can find broad applications in biology, medicine, microelectronics, and nanoplasmonics. For developing biomedical imaging applications, such matrices can be immersed in the liquid samples or cultures and micromanipulated inside such solutions.

## ACKNOWLEDGMENT

The authors thank I. Vitebskiy, J. Derov, and M. Schmitt for stimulating discussions. This work was supported by the U.S. Army Research Office (ARO) through Dr. J. T. Prater under Contract No. W911NF-09-1-0450 and by Center for Metamaterials, an NSF I/U CRC, Award No. 1068050. Also, this work was sponsored by the Air Force Research Laboratory (AFRL/RYD) through the AMMTIAC contract with Alion Science and Technology.